\documentclass[prb,superscriptaddress,citeautoscript,longbibliography,reprint]{revtex4-1}
\setcitestyle{numbers,square}
\usepackage[para]{threeparttable}
\usepackage[english]{babel}
\usepackage{array}
\usepackage{multirow}

\usepackage{pdftexcmds}
\usepackage[version=3]{mhchem}
\usepackage{siunitx}
\usepackage{xcolor}
\usepackage{graphicx}
\usepackage{dcolumn}
\usepackage{bm}
\usepackage{amsmath}
\usepackage{tikz}
\usepackage{pgfplots}
\usepackage{natbib}
\pgfplotsset{compat=1.16}

\usepackage{hyperref}
\hypersetup{
    colorlinks,%
    citecolor=blue,%
    linkcolor=blue,%
    urlcolor=blue
}

\newcommand{\orcid}[1]{\href{https://orcid.org/#1}{\includegraphics[width=8pt]{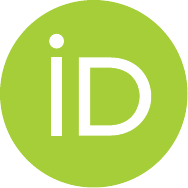}}}

\begin{document}

\title{Electronic properties of the Weyl semimetals Co$_2$MnX (X=Si, Ge, Sn)}

\author{Abhishek Sharan\orcid{0000-0002-6394-2199}}
\affiliation{Department of Physics and Astronomy, University of Delaware, Newark, DE 19716, USA}
\affiliation{Department of Materials Science and Engineering, University of Delaware, Newark, DE 19716, USA}

\author{Felipe Crasto de Lima\orcid{0000-0002-2937-2620}}
\affiliation{Department of Materials Science and Engineering, University of Delaware, Newark, DE 19716, USA}
\affiliation{Instituto de F\'isica, Universidade Federal de Uberlândia,
C.P. 593, 38400-902, Uberlândia, MG, Brazil}

\author{Shoaib Khalid\orcid{0000-0003-3806-3827}} 
\affiliation{Department of Physics and Astronomy, University of Delaware, Newark, DE 19716, USA}
\affiliation{Department of Materials Science and Engineering, University of Delaware, Newark, DE 19716, USA}

\author{Roberto H. Miwa\orcid{0000-0002-1237-1525}}
\affiliation{Instituto de F\'isica, Universidade Federal de Uberlândia,
C.P. 593, 38400-902, Uberlândia, MG, Brazil}

\author{Anderson Janotti\orcid{0000-0002-0358-2101}}
\affiliation{Department of Materials Science and Engineering, University of Delaware, Newark, DE 19716}

\date{\today}%

\begin{abstract}

Using first-principles electronic structure calculations, we show that ferromagnetic Heusler compounds Co$_2$MnX (X= Si, Ge, Sn) present non-trivial topological characteristics and belong to the category of Weyl semimetals. These materials exhibit two topologically interesting band crossings near the Fermi level. These band crossings have complex 3D geometries in the Brillouin zone and are characterized by non-trivial topology as Hopf links and chain-like nodal lines, that are protected by the perpendicular mirror planes. The spin-orbit interaction split these nodal lines into several zero-dimensional Weyl band crossings. Unlike previously known topologically non-trivial Heusler materials, these majority-spin band crossings lie in the band gap of minority spin bands, potentially facilitating its experimental realization.

\end{abstract}

\maketitle

\section{Introduction}

Since the discovery of Weyl semimetallic behavior in TaAs \cite{Weng2015, Lv2015, Sun2016, Xu2015}, several topological semimetals have been predicted, including nodal-line semimetals \cite{Fang2016}, \textit{Z}$_2$-topological semimetals \cite{khalid2019topological,nayak2017multiple,khalid2020}, non-symmorphic nodal-chain metals \cite{Bzdusek2016}, Kramers Weyl fermions \cite{Chang2018}, and magnetic Dirac semimetals \cite{Tang2016, Young2017, Wang2017}. Topological semimetals can further be classified and characterized by the dimensionality of band crossings in the bulk Brillouin zone (BZ). In Dirac or Weyl semimetals, conduction and valence bands cross at discrete points in the BZ, i.e., at zero dimensional crossings. For nodal-line semimetals, the band crossings follow a line in the BZ and hence, they are one dimensional; these band crossing may follow complex geometries in the BZ. 

Heusler compounds host a variety of non-trivial topological phases, including topological insulators (TI) and Weyl semimetals (WSM). In general, they offer high tunability of electronic properties, owing to the stability of a wide range of chemical combinations in different sites of their crystal structure \cite{Graf2011}. 
Rare-earth half-Heusler compounds PtLnBi (Ln = Y, La, and Lu) were one of the first topological materials predicted based on electronic structure calculations, with large spin-orbit coupling and strong band inversion \cite{Chadov2010, Xiao2010,Lin2010}, and later experimentally verified for PtLuBi, PtYBi \cite{Liu2016} and PtLuSb \cite{Logan2015}. Hopf-link semi-metallic phases have been predicted in magnetic Co$_2$TiX (X = Si, Ge and Sn) and Co$_2$MnGa full-Heuslers \cite{Chang2016, Chang2017}.  Slight changes in chemical compositions may lead to significant modifications in their electronic structure, potentially resulting in novel topological phases.

In this work we investigate non-trivial topological features in the full-Heusler compounds Co$_2$MnX (X= Si, Ge and Sn). These materials are ferromagnetic with high Curie temperatures \cite{Vanengen1983, Webster1971}. The majority spin bands near the Fermi level exhibit several three-dimensional band crossings in the bulk BZ. Some of these band crossings, which are protected by different crystal symmetries, are entangled and are classified as Hopf links \cite{Chang2017}. Upon breaking time reversal symmetry, these band crossings split up into several zero-dimensional crossings called Weyl nodes. These Weyl nodes exhibit topologically non-trivial characteristics that may be easier to probe experimentally due to the half-metallic character of these materials. 

\begin{figure}[h]
\centering
\includegraphics[width=0.9\columnwidth]{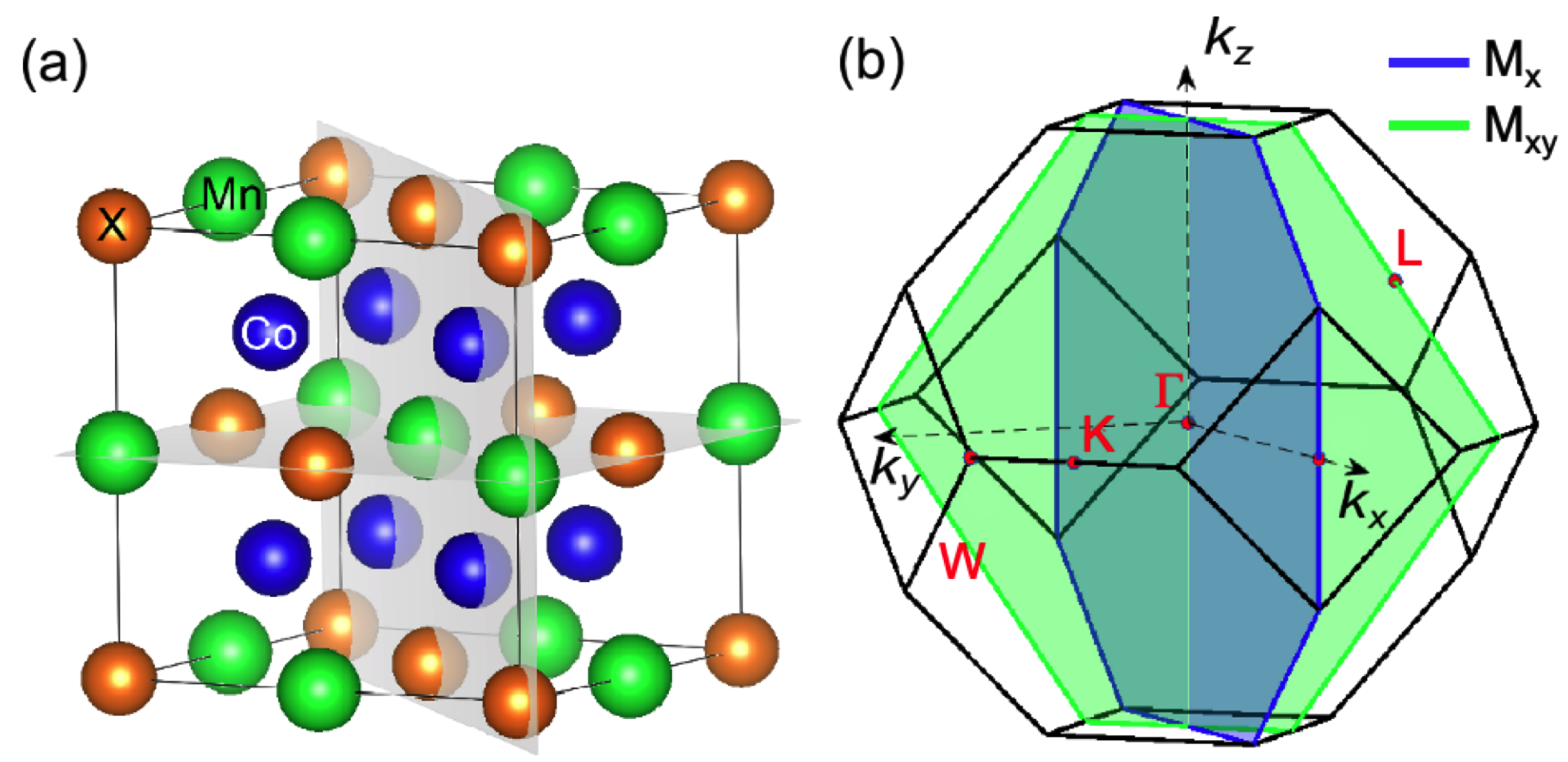}
\caption{\label{heusler:full_h_structure} (a) Full-Heusler crystal structure of Co$_2$MnX (X = Si, Ge or Sn), and (b) corresponding Brillouin zone. Two of the mirror-symmetry planes are shown in (a) and (b).
}
\end{figure}

\begin{figure*}[t]
\centering
\includegraphics[width=1.75\columnwidth]{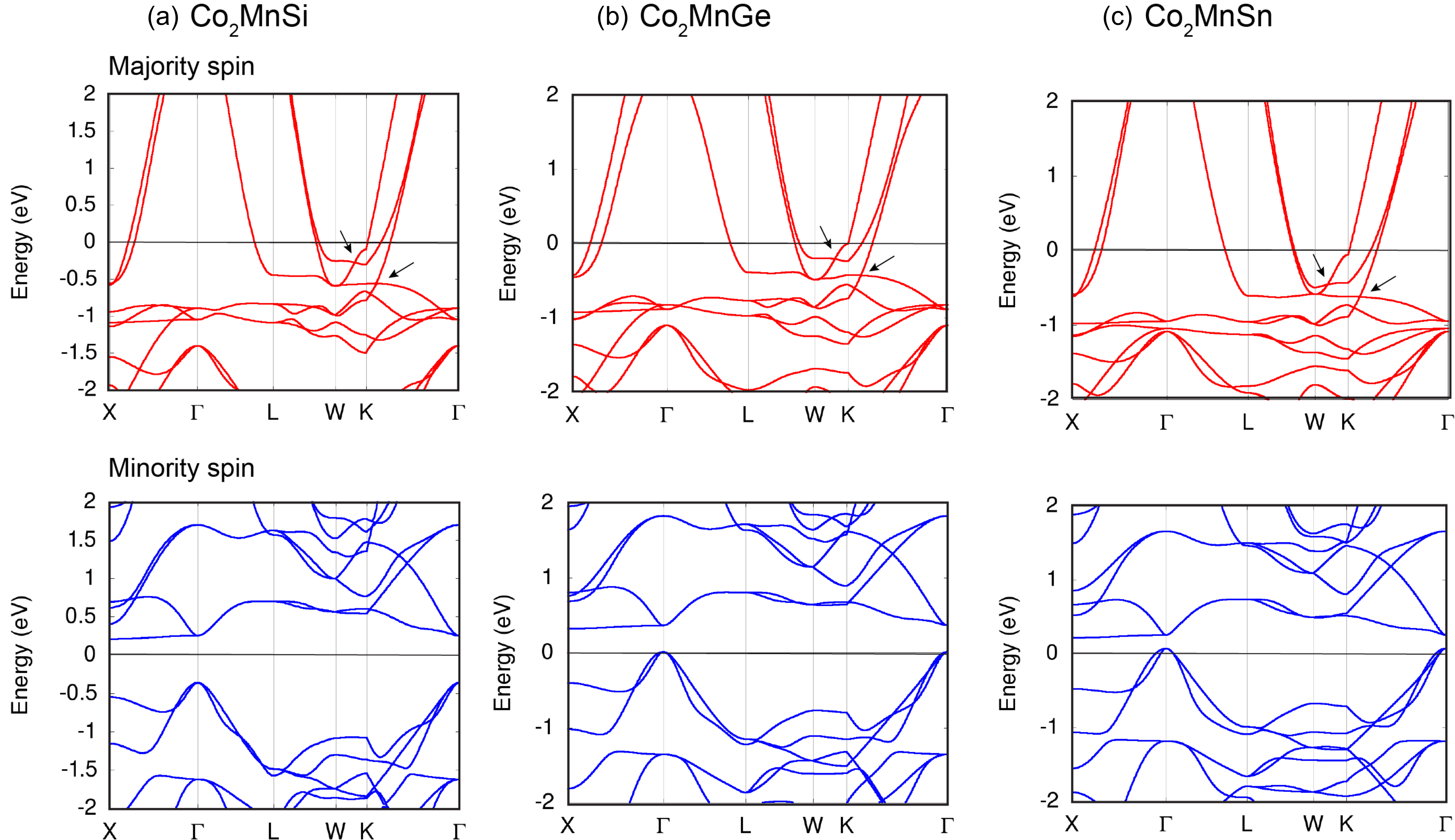}
\caption{\label{heusler:full_h_bandstr} Majority and minority spin band structures of (a) Co$_2$MnSi, (b) Co$_2$MnGe, and (c) Co$_2$MnSn. The zero in energy is taken as the Fermi level E$_{\rm{F}}$. Relevant band crossings near E$_{\rm{F}}$ in majority spin band structures are indicated by arrows.
}
\end{figure*}

\section{Computational Approach}

The calculations are based on the generalized Koh-Sham theory \cite{Hohenberg1964, Kohn1965} with generalized gradient approximation of Perdew-Burke-Ernzerhof revised for solids (PBEsol) \cite{Perdew2008}, as implemented in the VASP code \cite{Kresse1996a, Kresse1996c}. The interaction between the valence electrons and the ionic cores are treated using projector augmented wave potentials \cite{Blochl1994b, Kresse1999a}. The calculations are performed using a primitive cell of four atoms, with 350 eV energy cutoff for the plane-wave expansion and 16$\times$16$\times$16  $\Gamma$-centered mesh of $k$ points for integrations over the BZ. Tight-binding calculations are performed using the Wannier90 code \cite{Mostofi2014}, with hamiltonian extracted from the density functional theory calculations. Topological features of the band structure are computed with the tight-binding hamiltonian using the WannierTools code \cite{Wu2017}.

\section{Results and Discussion}

\subsection{Crystal Symmetry of Co$_2$MnX (X=Si, Ge, Sn)}
\label{full_h_crystal}

The full-Heusler crystal structure of Co$_2$MnX (X = Si, Ge or Sn), space group $Fm\overline{3}m$, is composed of four interpenetrating face centered cubic sub-lattices, as shown in Fig.~\ref{heusler:full_h_structure}(a). The corresponding BZ is shown in Fig.~\ref{heusler:full_h_structure}(b). The calculated lattice parameters of Co$_2$MnSi, Co$_2$MnGe and Co$_2$MnSn are 5.56 $\rm{\AA}$, 5.65 $\rm{\AA}$, and 5.90 $\rm{\AA}$ are in good agreement with the experimental values 5.66 $\rm{\AA}$ \cite{Cheng2001}, 5.75 $\rm{\AA}$ \cite{Cheng2001}, and 5.976 $\rm{\AA}$ \cite{Szytula1972}, respectively. The crystal structure has space inversion symmetry, with nine mirror-symmetry planes, $M_x (x=0)$, $M_y (y=0)$, $M_z (z=0)$, $M_{xy} (x=y)$, $M_{xz} (x=z)$, $M_{yz} (y=z)$, $M_{x\overline{y}} (x=-y)$, $M_{x\overline{z}} (x=-z)$ and $M_{y\overline{z}} (y=-z)$. These symmetry planes play important role in the topological property of these materials, as discussed in the subsequent sections.

\subsection{Electronic Band Structure of Co$_2$MnX (X=Si,Ge, and Sn)}

Co$_2$MnSi and Co$_2$MnGe have been characterized as half-metallic ferromagnets with magnetization $M$=5$\rm{\mu_B}$ per formula unit and Curie temperatures greater than 900 K \cite{Cheng2001}. These materials have 29 valence electrons per formula unit ($N_v$=29) and satisfy the Slater Pauling rule \cite{Fecher2006}, according to which $M=N_v-24$.
Here we find that Co$_2$MnSi is a half-metal with sizable gap of 0.72 eV in the minority-spin channel [Fig.~~\ref{heusler:full_h_bandstr}(a)], in agreement with previous electronic structure calculations \cite{Galanakis2002,Kandpal2007} and photoemission experiments \cite{Andrieu2016}. 

In the case of Co$_2$MnGe and Co$_2$MnSi, our calculations show that they are on the brink of becoming half-metals, with the Fermi level only touching the top of the valence band in the minority-spin channel of Co$_2$MnGe [Fig.~\ref{heusler:full_h_bandstr}(b)], with a band gap of 0.35 eV in agreement with previous calculations \cite{Picozzi2002}.  We note that recent angle resolved photoemission spectorscopy (ARPES) measurements in Co$_2$MnGe find the Fermi level at less than 0.1 eV above the top of the valence band in the minority spin gap \cite{Kono2020}, in overall good agreement with our calculations. However, we note that ARPES is highly surface sensitive, and previous measurements on Co$_2$MnSi indicate a strong dependence of the Fermi level position on the surface termination \cite{Andrieu2016}. 
For Co$_2$MnSn, the Fermi level is slightly below the top of the valence band (with a small band gap of 0.18 eV) leading to a very small, yet non zero, minority-spin density of states at the Fermi level [Fig.~\ref{heusler:full_h_bandstr}(c)], also in agreement with previous calculations \cite{Picozzi2002}.  

Regarding the topological features, there are two interesting band crossings in the majority-spin band structure near the Fermi level.  It is worth noting that here, different from previously explored Co$_2$MnGa \cite{Chang2017}, the half-metal-like behavior of Co$_2$MnX (X=Si, Ge, and Sn), with very small minority-spin density of states in the Ge and Sn-based compounds, may facilitate the experimental observations of Weyl nodes. Since the nature of the band crossings in the majority-spin channel is the same in these three compounds, as shown  in Fig.~\ref{heusler:full_h_bandstr}, we opt to focus the discussion on the case of Co$_2$MnGe. The conclusions for the other two materials are similar.

\begin{figure*}
\centering
\includegraphics[width=1.75\columnwidth]{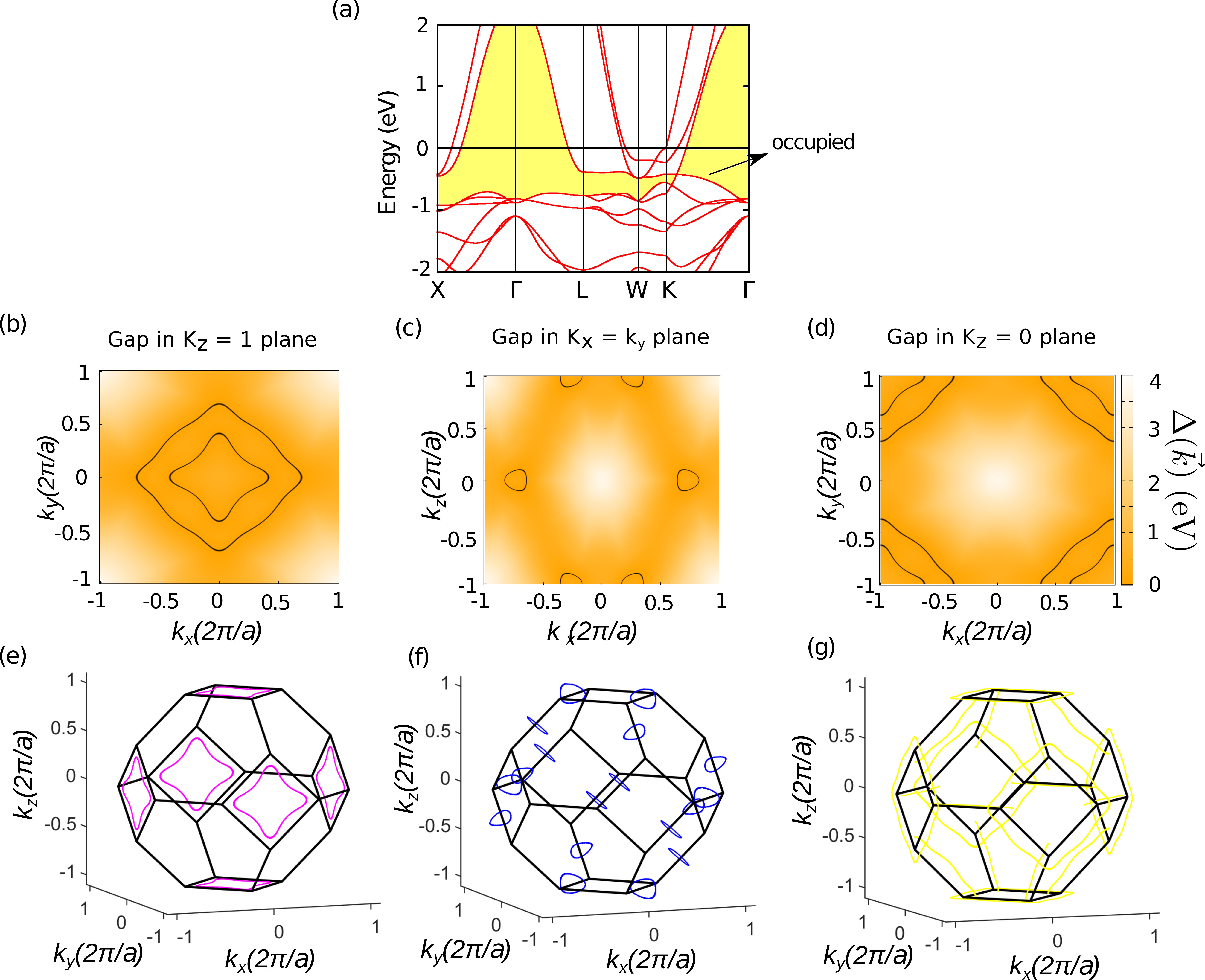}
\caption{Tight-binding band structure Co$_2$MnGe.  (a) Majority-spin band structure with the shaded region indicating the gap between the highest occupied and lowest unoccupied band. Projection of the energy gap, $\Delta(\vec{k})$, between highest occupied and lowest unoccupied band on different planes, (b) k$_z$ = 1 plane, (c) k$_x$ = k$_y$ plane, and (d) k$_z$ = 0, showing three different types of nodal lines. The location of the nodal lines in the 3D  Brillouin zone are shown in (e), (f) and (g).
}
\label{heusler:nodal_line_1}
\end{figure*}

\begin{figure}[h]
\centering
\includegraphics[width=0.8\columnwidth]{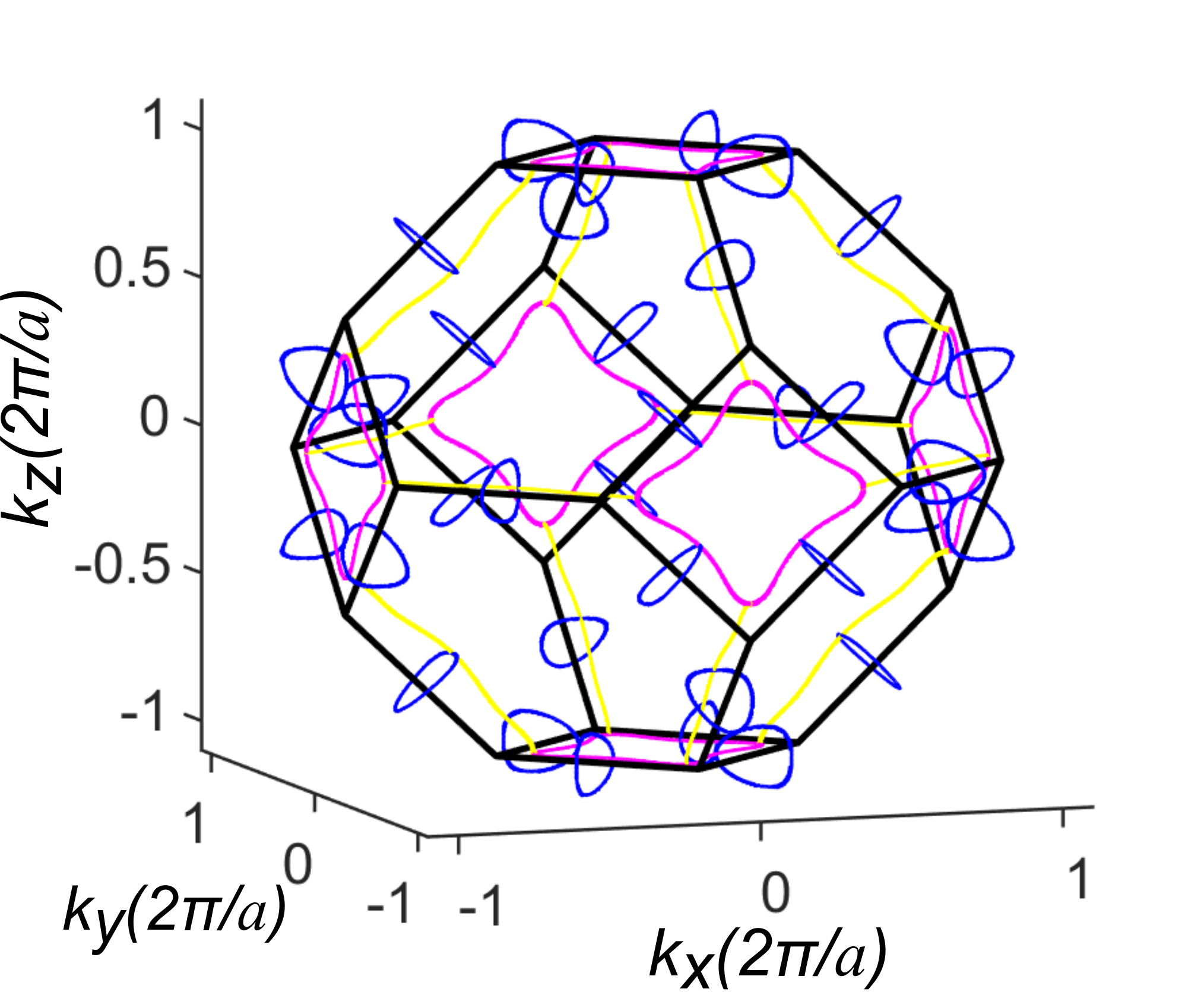}
\caption{Connection of the different nodal lines from Fig.~\ref{heusler:nodal_line_1}(e)-(g), in the 3D Brillouin zone of Co$_2$MnGe.
}
\label{heusler:nodal_line_1_all}
\end{figure}

\subsection{Nodal lines in Co$_2$MnGe}

The two topologically interesting band crossings near the Fermi level in the majority-spin band structure are indicated in Figs.~\ref{heusler:full_h_bandstr}(a)-(c). To understand the geometry of these band crossings in the three-dimensional BZ we track them separately. Figure~\ref{heusler:nodal_line_1}(a) shows the majority-spin band structure of Co$_2$MnGe obtained from a Wannier base tight-binding (TB) model. The TB hoping parameters were extracted from first-principles calculations,which presents very good agreement with DFT electronic structure.

To track the crossing between two bands we have calculated the energy gap separation of the pair of bands for each k-point, $\Delta(\vec{k}) = E_n(\vec{k}) - E_{n-1}(\vec{k})$, along different mirror symmetry planes. First, for the lower band crossing in the majority-spin of Fig.~\ref{heusler:full_h_bandstr}(b), we populate the band indicated by the arrow in Fig.~\ref{heusler:nodal_line_1}(a) and find the gap between this highest occupied and the next, lowest unoccupied, band [shaded region in Fig.~\ref{heusler:nodal_line_1}(a)]. The projection of the gap between these two bands is shown in Fig.~\ref{heusler:nodal_line_1}(b)-(d) along $\rm{k_z = 1}$, $\rm{k_x=k_y}$ and $\rm{k_z=0}$ plane, which are also mirror symmetry planes as discussed in section \ref{full_h_crystal}.

\begin{figure*}
\centering
\includegraphics[width=1.4\columnwidth]{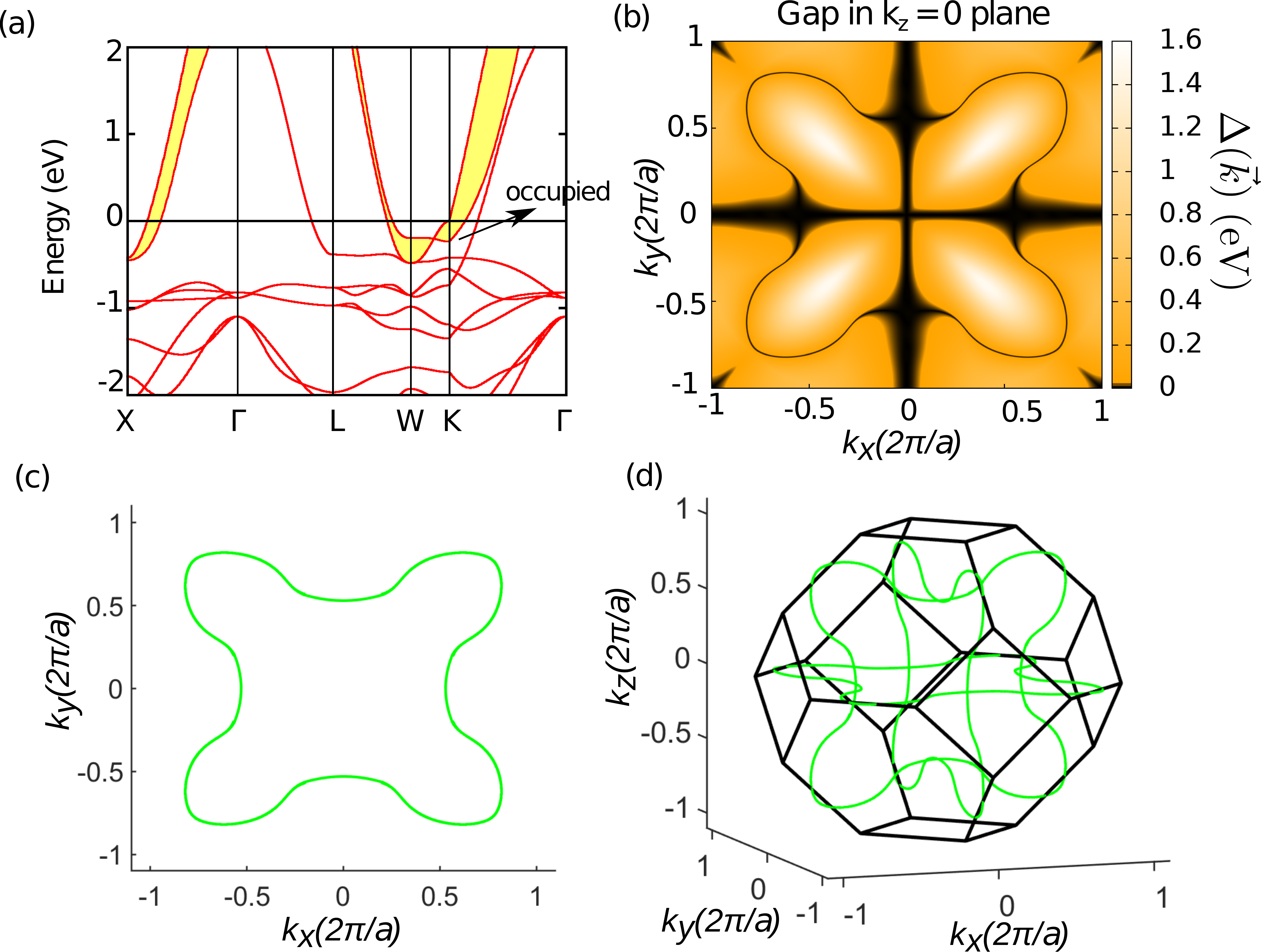}
\caption{Band structure of Co$_2$MnGe highlighting the higher band crossing near the Fermi level.  (a) Majority-spin band structure with the arrow indicating highest occupied band in the TB model. (b) The gap ($\Delta(\vec{k})$) between the highest occupied and lowest unoccupied bands [shaded region in (a)] projected on the $\rm{k_z = 0}$ plane. (c) The nodal line shown in k$_x$-k$_y$ plane. (d) The dispersion of this nodal line in 3D Brillouin zone.
}
\label{heusler:nodal_line_2}
\end{figure*}

There are three different types of nodal lines that can be observed here. The dispersion of these nodal lines in the BZ is shown in Fig.~\ref{heusler:nodal_line_1}(e)-(g). The pink and yellow nodal lines are protected by $M_x$, $M_y$ and $M_z$ mirror planes, while blue nodal lines are protected by $M_{xy}$, $M_{xz}$ and $M_{yz}$ mirror planes. Interestingly, all these three types of nodal lines are entangled, forming a nodal chain \cite{Bzdusek2016, Chang2017}, as shown in Fig.~\ref{heusler:nodal_line_1_all}.

Two nodal lines (in yellow) are interlinked as shown in Fig.~\ref{heusler:nodal_line_1}(g) and Fig.~\ref{heusler:nodal_line_1_all}, forming a Hopf link \cite{Chang2017}. Each nodal line in pink is connected to four yellow nodal lines and four blue nodal lines. The trajectory of these nodal lines in the BZ, protected by different mirror symmetries, gives rise to another category of topological semimetals. Similar network of nodal lines has also been found in Co$_2$MnGa \cite{Chang2017}, due to similarity in the nature of associated bands.

Next, we track the upper band crossing near the Fermi level, along the WK direction, of the majority-spin band structure shown in Fig.~\ref{heusler:full_h_bandstr}(b), using a similar approach as described above.  The bands up to the crossing point in Fig.~\ref{heusler:nodal_line_2}(a) are occupied and the gap between this highest occupied and lowest unoccupied band is computed and projected along different mirror symmetry planes. We observe that these bands cross along $\rm{k_x = 0}$, $\rm{k_y = 0}$ and $\rm{k_z = 0}$ planes. The projection of the band gap in $\rm{k_z = 0}$ is shown in Fig.~\ref{heusler:nodal_line_2}(b) and (c). The dispersion of this nodal line in the BZ is shown in Fig.~\ref{heusler:nodal_line_2}(d). This kind of nodal lines are also seen in the Heusler Weyl semimetals Co$_2$TiX (X = Si, Ge or Sn) \cite{Chang2016}.

\subsection{Effects of Spin-orbit Coupling on the electronic structure of Co$_2$MnGe}

Here we analyze the effects of spin-orbit coupling on the band structure of Co$_2$MnX (X=Si, Ge, Sn), again, taking Co$_2$MnGe as example. Since these materials are ferromagnetic, the symmetry of the system and electronic properties will depend on the magnetization direction. We calculate the magnetic anisotropy energy for magnetization along [001], [110], and [111] directions. The energy of the system along these three magnetization directions are very similar, differing by less than 0.1 meV per formula unit. Adding spin-orbit coupling, mirror symmetry is broken along certain planes. For instance, if magnetization is along [001], the $M_z (k_z = 0)$ mirror symmetry is preserved and the nodal line along this plane is preserved, while nodal lines along the other mirror planes are gapped as the respective mirror symmetry is broken, resulting in the emergence of  Weyl nodes. Due to the small magnetization anisotropy energy, the magnetic orientation can be tuned by an external field, and thus the formation and location of the Weyl nodes can be controlled \cite{Chang2016, NATCOMMli2020}.

\begin{figure}
\centering
\includegraphics[width=\columnwidth]{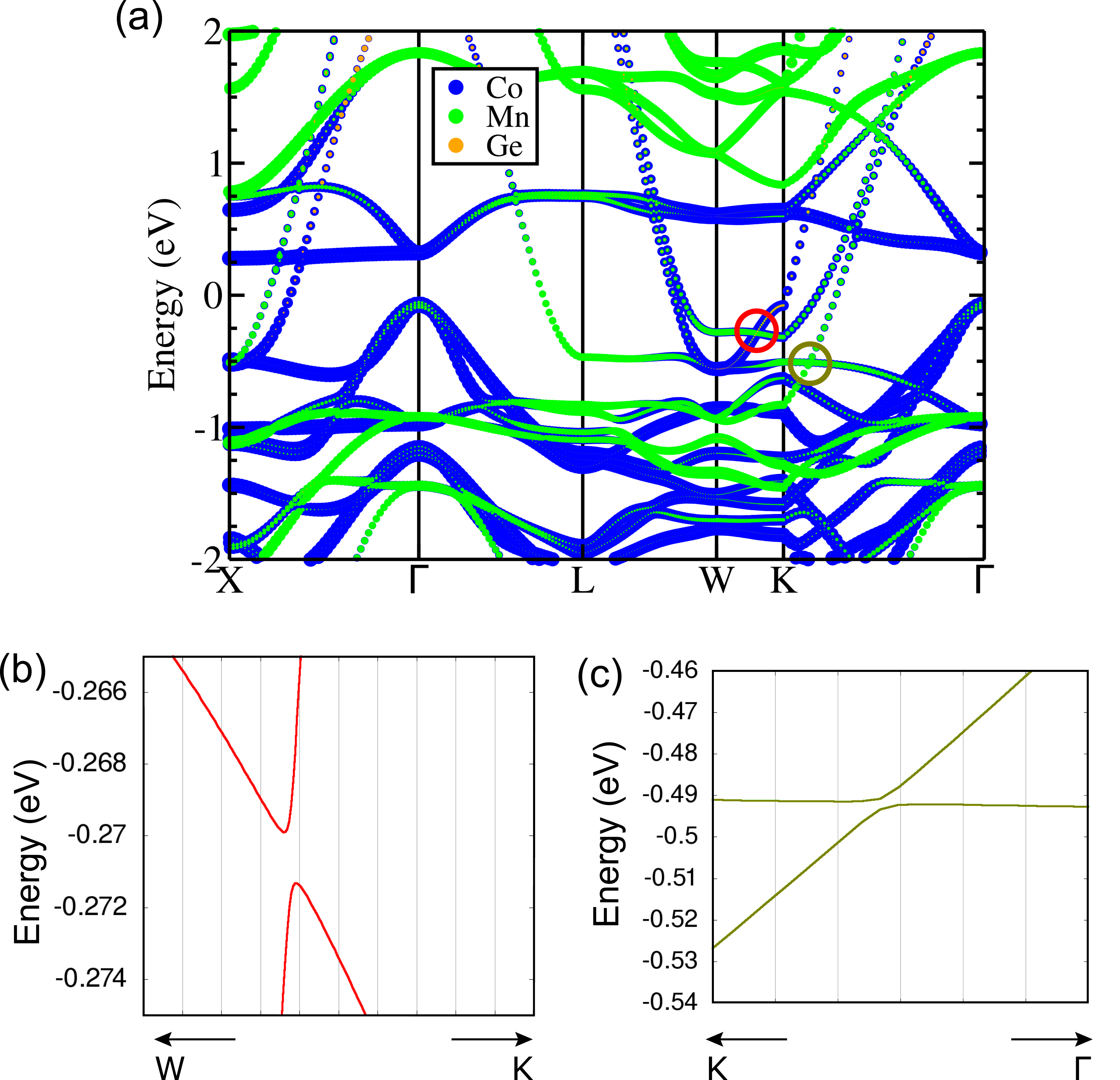}
\caption{\label{heusler:spin_orbit_bandstr} (a) Atomic projected electronic structure of Co$_2$MnGe in presence of spin-orbit coupling and magnetization along [111] direction. The band crossings discussed in this section are highlighted by red and bronw circles and enlarged in (b) and (c).
}
\end{figure}
 
\begin{figure}[h]
\centering
\includegraphics[width=\columnwidth]{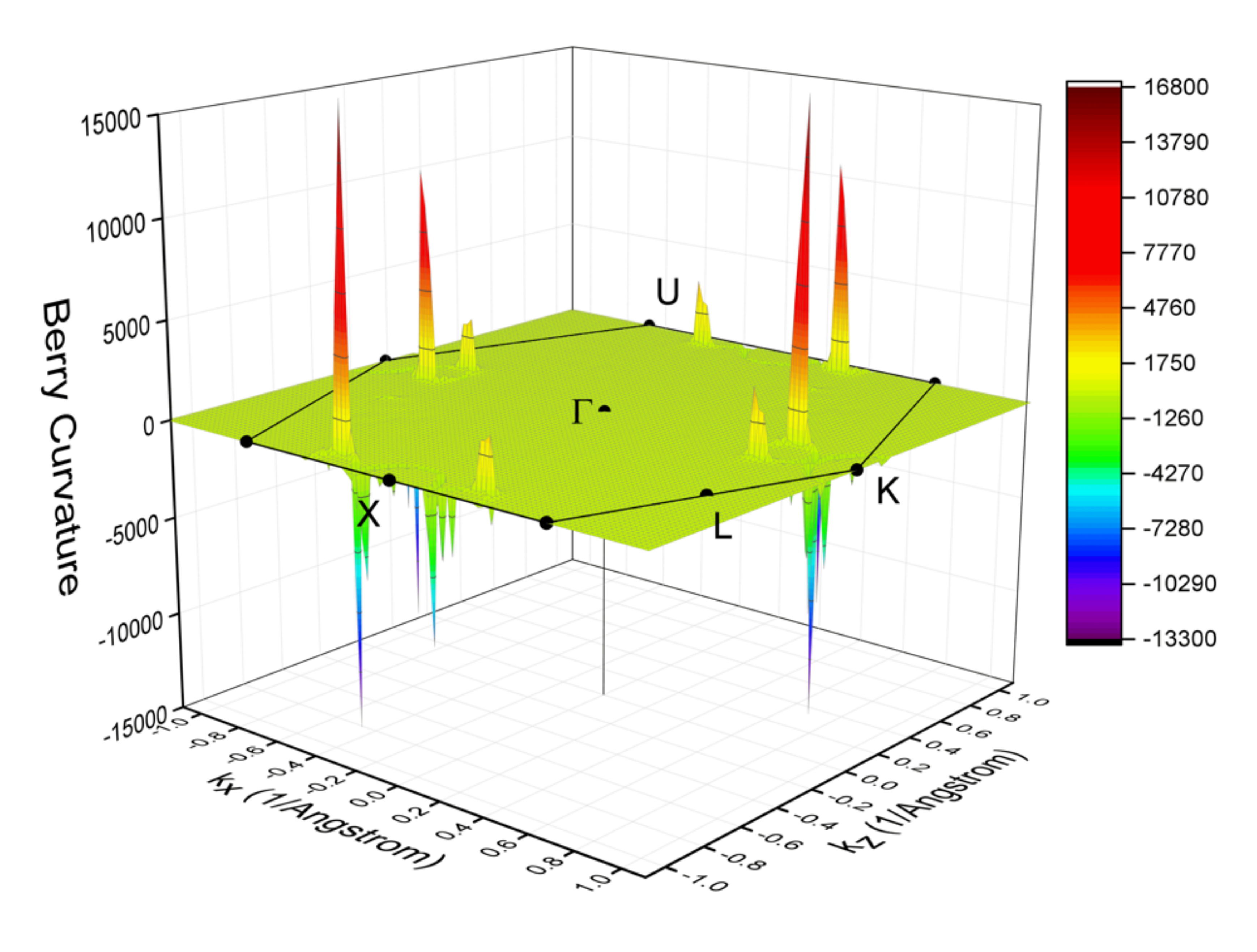}
\caption{Variation of Berry curvature for Co$_2$MnGe in the $k_y=0$ plane. 
}
\label{heusler:berry_curvature}
\end{figure}

Figure~\ref{heusler:spin_orbit_bandstr}(a) shows the band structure of Co$_2$MnGe with inclusion of spin-orbit coupling and magnetization along the [111] direction;  Figure~\ref{heusler:spin_orbit_bandstr}(b) and (c) show enlarged views around the two band crossings near the Fermi level. Here we can see that the Ge orbitals do not contribute significantly to the crossing bands, being dominated by the Co and Mn orbitals. As expected, given the spin-orbit contribution, the band crossings are gapped. However, the effect of spin-orbit coupling on this gap is very small, with band splitting of less than 5 meV, and and does not depend on the anion. This can be explained by the very small contribution from orbitals of the X atom to the bands near the crossing. As a result of mirror symmetry breaking, Weyl nodes are expected to appear. Weyl nodes of opposite chiralities are source and sink of Berry curvature. The Berry curvature is calculated using the Kubo formula \cite{Xiao2010}:
\begin{equation}
\label{berry_cur}
\Omega_{n}^{\gamma} = 2i\hbar^2 \sum_{m \neq n} \frac{\langle u_n(\vec{k})|\hat{v_{\alpha}}|u_m(\vec{k})\rangle \langle u_m(\vec{k})|\hat{v_{\beta}}|u_n(\vec{k})\rangle}{(E_n(\vec{k}) - E_m(\vec{k}))^2},
\end{equation}
where $\Omega_{n}^{\gamma}$ is the Berry curvature for band $n$, $\hat{v}_{\alpha} = \frac{1}{\hbar}\frac{\partial \hat{H}}{\partial k_{\alpha (\beta, \gamma)}}$ is the velocity operator for $\alpha, \beta, \gamma = x, y, z$ and $|u_n(\vec{k}) \rangle$ and $E_n(\vec{k})$ are the eigenvectors and eigenvalues of the Hamiltonian $\hat{H}(\vec{k})$ respectively. For the computation of Berry curvature we have taken the summation in $\Omega_{n}^{\gamma}$ over all bands up to the lower band crossing [brown circle in Fig.~\ref{heusler:spin_orbit_bandstr}(a)]. The plot of Berry curvature in $k_y$=0 plane is shown in Fig.~ \ref{heusler:berry_curvature}. Several discontinuities are observed in the plot of Berry curvature along in the $k_y=0$ plane, where Weyl nodes appear. High positive (negative) spikes in the values of Berry curvature is associated with points of positive (negative) chirality, representing the source (sink) of Berry curvature. Several pairs of Weyl nodes also appear along other symmetry planes. The separation between Weyl nodes of opposite chirality in momentum space is very small ($\sim$0.02 in units of 2$\pi$/$a$). This makes it quite challenging to probe the existence of these Weyl points and to observe the surface Fermi arcs. 

\begin{figure}[h]
\centering
\includegraphics[width=0.8\columnwidth]{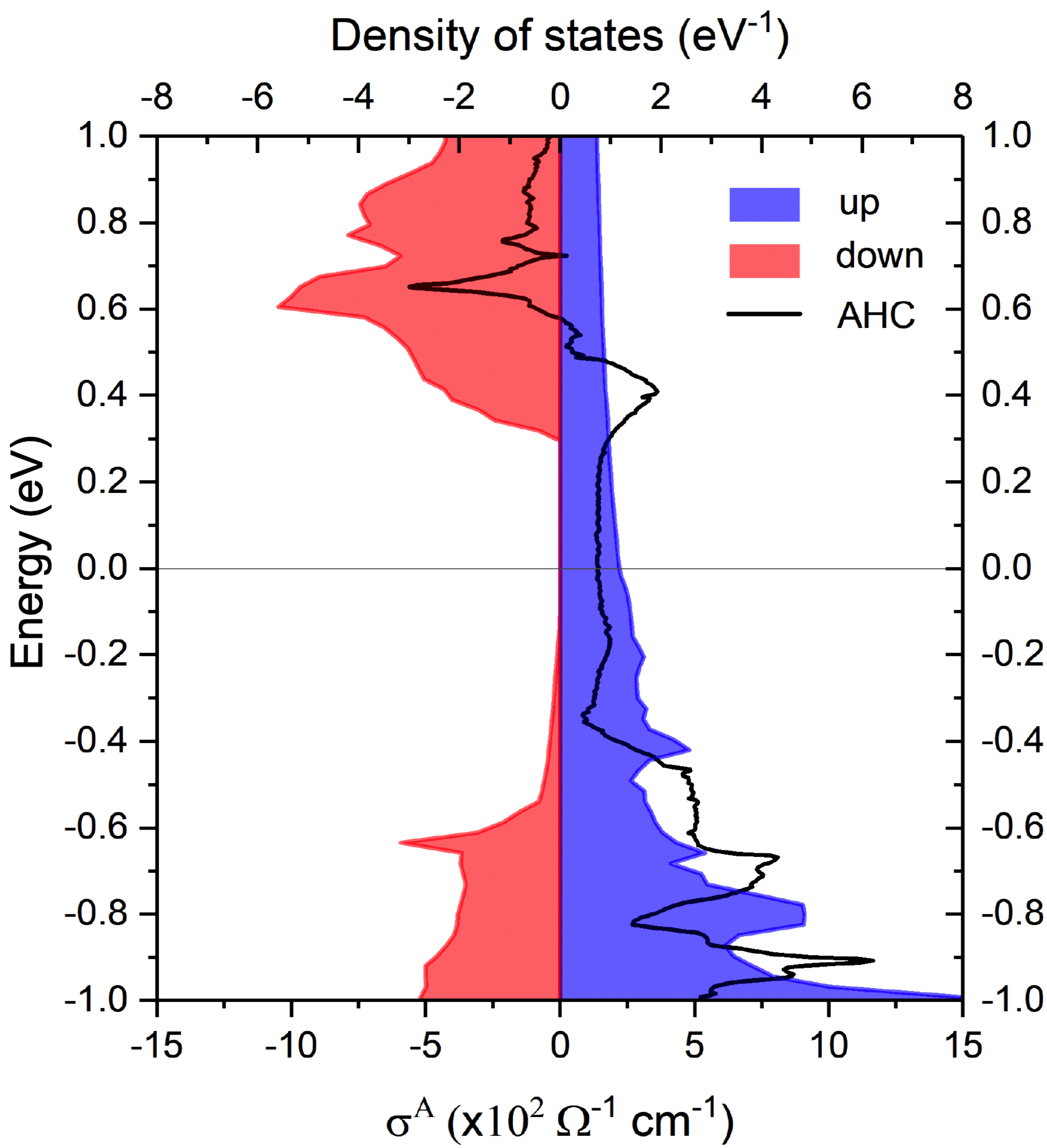}
\caption{Anomalous Hall conductivity ($\sigma^A$) and spin resolved density of states for Co$_2$MnGe. 
}
\label{heusler:ahc}
\end{figure}

\subsection{Anomalous Hall Conductivity of Co$_2$MnGe}

Heusler materials exhibiting Weyl points typically show large anomalous Hall conductivity (AHC). The value of AHC depends on the proximity of Weyl points to the Fermi level and the separation between Weyl nodes of opposite chirality in momentum space. A range of AHC values have been reported for Heusler materials: Co$_2$TiSn shows AHC of $\sim$100 $\rm{\Omega^{-1}cm^{-1}}$ while Co$_2$MnAl shows high AHC of $\sim$2000 $\rm{\Omega^{-1}cm^{-1}}$ \cite{Kubler2012}. The intrinsic AHC conductivity $\sigma_{xy}$ is calculated using:
\begin{equation}
\sigma_{xy} = -\frac{e^2}{\hbar} \int_{BZ} \frac{d^3\vec{k}}{(2\pi)^3}f_n(\vec{k})\Omega^{z}(\vec{k}),
\end{equation}
where, $\Omega^{z}(\vec{k})$ is the Berry curvature computed using Eq.~\ref{berry_cur} and $f_n(\vec{k})$ is the Fermi-Dirac distribution function. The plot of AHC overlayed on top of the spin-resolved density of states for Co$_2$MnGe is shown in Fig.~\ref{heusler:ahc}. We predict AHC of 140 $\rm{\Omega^{-1}cm^{-1}}$, which falls in between the AHC calculated for Co$_2$MnSi (228 $\rm{\Omega^{-1}cm^{-1}}$) and Co$_2$MnSn (118 $\rm{\Omega^{-1}cm^{-1}}$) \cite{Kubler2012}. Given the similarities in the electronic structure of the Co$_2$MnX compounds and the SOC effect on the band crossings having the major contributions from Co and Mn orbitals, a difference in AHC for these three compounds is expected to depend on the relative energetic position of the node-line crossing. For instance, from the AHC plot we see a large conductivity of $0.6$\,eV below the Fermi energy in Co$_2$MnGe; on the other hand for Co$_2$MnAl the closeness of the crossing to the Fermi energy dictates its much higher AHC \cite{NATCOMMli2020}.

\section{Summary and Conclusion}

We studied the family of Heusler compounds Co$_2$MnX (X = Si, Ge or Sn) as magnetic topological Weyl semimetals. These materials are ferromagnetic with a magnetic moment $M$=5 $\mu_B$ per formula unit. Through spin-resolved band structure, we show that these materials exhibit two interesting band crossings in the majority-spin channel near the Fermi level. Using Co$_2$MnGe as an example, we discuss the nodal line features corresponding to the two band crossings near the Fermi level. Several nodal lines are shown to be present in the system which are protected by different mirror-symmetry planes. These nodal lines show three-dimensional features, some of which are entangled. In presence of spin-orbit coupling and finite magnetization in the system, mirror symmetry is preserved only in the plane perpendicular to the magnetization direction and is broken along other mirror planes. Consequently, the nodal lines are gapped and Weyl points emerge. We show the presence of Weyl nodes of opposite chiralities in Co$_2$MnGe, which are the source and sink of Berry curvature. We also predict Anomalous Hall conductivity of 140 $\rm{\Omega^{-1}cm^{-1}}$ for Co$_2$MnGe.

\section{Acknowledgements}
This work was supported by the U.S. Department of Energy under Award No.~DE-SC0014388, and  the computational resources of the National Energy Research Scientific Computing Center (NERSC), a U.S. Department of Energy Office of Science User Facility operated under Contract No. DE-AC02-05CH11231. F.~C.~L. and R.~H.~M. also acknowledge the Brazilian agencies FAPEMIG, and CNPq, and the computing facilities LNCC (SCAFMat2) and CENAPAD-SP.

\bibliography{wsm}

\end{document}